\documentclass[prl,amsmath,amsfonts,amssymb,letterpaper,superscriptaddress,twocolumn,showpacs%
]{revtex4}
\usepackage{bm}
\usepackage{graphicx}

\newcommand\arrow\vec

\def\vec#1{\bm{#1}}

\def\negspace{\!}

\def\lrsub#1#2#3{{\vphantom{#1}}_{#2} \negspace {#1} \negspace {\vphantom{#1}}_{#3}}

\def\bra#1{\left\langle {#1} \right\rvert}
\def\ket#1{\left\lvert {#1} \right\rangle}

\def\inprod#1#2{\left\langle {#1} | {#2} \right\rangle}

\def\inprodsubsub#1#2#3#4{\lrsub {\inprod{#1}{#2}} {#3} {#4}}

\def\pqinprod#1#2{\inprodsubsub{#1}{#2} p q}
\def\qpinprod#1#2{\inprodsubsub{#1}{#2} q p}
\def\pinprod#1#2{\inprodsubsub{#1}{#2} p p}
\def\qinprod#1#2{\inprodsubsub{#1}{#2} q q}
\def\pqbraket\pqinprod
\def\qpbraket\qpinprod
\def\pbraket\pinprod
\def\qbraket\qinprod

\def\outprod#1#2{\ket {#1}\!\bra {#2}}

\def\1{I}

\def\v0{{\bvec 0}}

\def\controlled#1{\mathrm{C}_{#1}}
\def\CZ{\controlled Z}

\begin{document}

\title{Ultracompact Generation of Continuous-Variable Cluster States}

\author{Nicolas C. Menicucci}
\email{nmen@princeton.edu}
\affiliation{Department of Physics, Princeton University, Princeton, NJ 08544, USA}
\affiliation{Department of Physics, The University of Queensland, Brisbane, Queensland 4072, Australia}

\author{Steven T. Flammia}
\affiliation{Department of Physics and Astronomy, University of New Mexico, Albuquerque, NM, 87131, USA}

\author{Hussain Zaidi}
\author{Olivier Pfister}
\email{opfister@virginia.edu}
\affiliation{Department of Physics, University of Virginia, 382 McCormick Road, Charlottesville, Virginia 22904-4714, USA}


\date{March~12, 2007}

\begin{abstract}
We propose an experimental scheme that has the potential for large-scale realization of continuous-variable (CV) cluster states for universal quantum computation.  We do this by mapping CV cluster-state graphs onto two-mode squeezing graphs, which can be engineered into a single optical parametric oscillator (OPO).  The desired CV cluster state is produced directly from a joint squeezing operation on the vacuum using a multi-frequency pump beam.  This method has potential for ultracompact experimental implementation.  As an illustration, we detail an experimental proposal for creating a four-mode square CV cluster state with a single OPO.
\end{abstract}

\maketitle

One-way quantum computation~\cite{Raussendorf2001} is a promising form of quantum computing (QC). Unitary gates are implemented by performing single-qubit measurements on a highly entangled ``cluster state''~\cite{Briegel2001}. Such states are particular cases of graph states~\cite{Hein2006} and are characterized by this ability to coherently manipulate quatum information using only local measurements and classical feedforward of measurement results.  One often discusses QC in terms of two-dimensional systems (qubits) but continuous quantum variables can also be used~\cite{Lloyd1999,Bartlett2002}. Continuous-variable (CV) cluster states~\cite{Zhang2006,Menicucci2006} generalize the concept of cluster states to quantum systems with continuous degrees of freedom by use of the quadratures, amplitude $q=(a+a^\dagger)$ and phase $p=-i(a-a^\dagger)$, of a boson field~\cite{Walls1994}. The preparation procedure exactly mirrors that for qubit cluster states, using the correspondence between the Pauli and continuous Weyl-Heisenberg groups~\cite{Lloyd1999}, and can be described in an analogous way to the graph state formalism~\cite{Hein2006} for qubit cluster states:  First, prepare each mode (represented by a graph vertex) in a phase-squeezed state, approximating a zero-phase eigenstate (analog of Pauli-$X$ eigenstates).  Then, apply a quantum nondemolition (QND) interaction of the form $\exp(i q_j q_k)$ (analog of controlled-phase $\CZ$) to each pair of modes $(j,k)$ linked by an edge in the graph. All QND gates commute (as do $\CZ$ gates), so the full multimode QND operator to be applied is $\exp(\tfrac i 2 \vec q{}^TA \vec q)$, where $\vec q = (q_1, \dotsc, q_N)^T$ is a vector of amplitude quadrature operators, and $A$ is the adjacency matrix for the graph. The resulting cluster state satisfies, in the limit of infinite squeezing, the relation~\cite{vanLoock2006a}
\begin{align}
\label{eq:CVCSdef}
	\vec p - A \vec q \to 0\;,
\end{align}
as a simultaneous zero-eigenstate of the $N$ components of $\vec p - A \vec q$.  Being infinitely squeezed, this state is unphysical (it has infinite energy) but we can approximate its Wigner function by any Gaussian strongly squeezed along the indicated quadratures.  Defining a ``CV cluster state'' as any of these approximating Gaussians is identical to the definition used in Ref.~\cite{vanLoock2006a} and more general than that studied in Refs.~\cite{Zhang2006,Menicucci2006}.  While only Gaussian operations are needed to create CV cluster states from the vacuum, using them for universal QC requires that at least one single-mode non-Gaussian measurement (such as photon-number resolving detection) be available~\cite{Menicucci2006}.

Although convenient theoretically, the above procedure is not optimal for experimental implementation because the QND gates contain in-line squeezers. It can be spectacularly simplified~\cite{vanLoock2006a} by use of the Bloch-Messiah reduction~\cite{Braunstein2005}, which transforms any Gaussian operation into the canonical form of a set of single-mode squeezers (e.g.,~optical parametric oscillators---OPOs) sandwiched between two multimode interferometers. With a vacuum input, the initial interferometer is irrelevant and any Gaussian CV $N$-mode cluster can be formed by $N$ single-mode vacuum squeezers (easier to implement than in-line ones) followed by a network of $O(N^2)$ beam splitters, i.e.\ a quadratically large, stable interferometer~\cite{vanLoock2006a}. Recently, a four-mode linear cluster state was demonstrated~\cite{Su2007}. 

In this Letter, we show that it is, in fact, possible to integrate all single-mode OPOs into {\em one multimode OPO}, pumped by an $O(N^2)$-mode field, and to eliminate the beam splitter network completely. This is equally resource-efficient as the proposal in Ref.~\cite{vanLoock2006a} but the complexity has been shifted from a stabilized $O(N^2)$-element interferometer (unwieldy for large $N$) to the nonlinear medium of a single OPO and the frequency content of the pump beam.  This scheme is much more compact and it has interesting prospects for scalability because it effectively represents the quantum entangled version of an optical frequency comb: as is well known, a femtosecond laser effectively compactifies $\sim 10^5$ phase-locked  continuous-wave lasers into a single beam~\cite{Haensch2006,Hall2006}. One of us showed that such a comb can be transformed into a GHZ state $\int dx\,\ket x_1\cdots\ket x_N$, where the subscripts denote consecutive comb lines, using a complete network of concurrent nonlinear interactions~\cite{Pfister2004,Bradley2005,Pfister2007}, and the nonlinear medium required to create four-mode entanglement in a single OPO has already been demonstrated~\cite{Pooser2005}. Engineering concurrent nonlinear interactions between an arbitrary number of modes is a complicated problem but it is now solvable in the general case by use of generalized quasi-phase-matching in photonic quasicrystals~\cite{Lifshitz2005}.  This enables arbitrarily difficult nonlinear interactions (e.g.,~simultaneous generation of the second, third, and fourth optical harmonics, all in different directions~\cite{Lifshitz2005}) to be engineered in a single OPO.

The central result of this paper is a mathematical connection between CV cluster states and two-mode squeezing (TMS) graph states~\cite{Pfister2007}.  CV cluster-state graphs have vertices representing phase-squeezed states and edges corresponding to QND {\em unitary\/} interactions and are the ones we wish to implement for one-way QC~\cite{Menicucci2006}.  TMS graphs have vertices representing vacuum inputs and edges denoting individual terms in the multimode squeezing {\em Hamiltonian\/} 
\begin{align}
\label{eq:HTMS}
	\mathcal H =-\frac i 2 \sum_{m,n} G_{mn}(a_m^\dag a_n^\dag - a_m a_n),
\end{align}
where $G$ denotes the adjacency matrix of the graph.  We prove that the two are related: any CV cluster state with a bipartite graph can be created by applying a single multimode squeezing Hamiltonian of the form of Eq.~\eqref{eq:HTMS} and any such Hamiltonian generates some CV cluster state.  We detail how to create a square cluster using this method with current technology.


Given a target CV cluster state, our goal is to effect a transformation on the quadrature operators such that Eq.~(\ref{eq:CVCSdef}) holds for the new quadratures.  We first collect $\vec q$ and $\vec p$ into a column vector $\vec x = (q_1, \dotsc, q_N, p_1, \dotsc, p_N)^T$.  Gaussian transformations on the vacuum in Hilbert space correspond to symplectic linear transformations on this vector in the Heisenberg picture~\cite{Simon1994}.  We denote by~$U_\alpha$ the symplectic transformation corresponding to a unitary that creates a CV cluster state from the vacuum.  The level of overall squeezing is represented by~$\alpha>0$, which should be as large as possible.  From Eq.~(\ref{eq:CVCSdef}) we have
\begin{align}
\label{eq:CVCStransform}
	\begin{pmatrix}
		-A & \1
	\end{pmatrix}
	U_\alpha \vec x_0 \to 0\;,
\end{align}
where the block matrix above is $N \times 2N$, $U_\alpha$ is $2N \times 2N$, and $\vec x_0$ is the vector of quadrature operators representing the vacuum state.  The arrow denotes 
the limit $\alpha \to \infty$. 

We have some additional freedom in Eq.~\eqref{eq:CVCStransform}.  After the transformation~$U_\alpha$ is applied, we can perform arbitrary phase shifts for each individual mode at the output, which we will represent with the matrix~$T$.  This is a passive transformation on the state, which can be effected simply by reinterpreting the output modes (i.e., no change to the physical apparatus used to create the state is required).  Therefore, we have that
\begin{align}
\label{eq:CVCSsufficient}
	\begin{pmatrix}
		-A & \1
	\end{pmatrix}
	T U_\alpha \to 0
\end{align}
is sufficient to conclude that $U_\alpha$ can be used to create a CV cluster state with adjacency matrix~$A$ from the vacuum.

As we will now show, if $A$ represents a bipartite graph, we can always do this with a multimode squeezing Hamiltonian.  
By definition, the nodes of a bipartite graph are partitioned into two sets such that 
all graph edges link one set to the other.
These graphs are also known as {\em two-colorable\/} graphs because the two sets (and the nodes each contains) can be assigned different colors.  Bipartite graphs include the square lattice graph of arbitrary size, which is universal for QC, and any of its subgraphs.  Star graphs (of any size) are also bipartite, with the node at the center being one color and the rest a different color.  As a counterexample, the triangle graph (and, more generally, any graph with an odd cycle in it) is not bipartite.

Consider a multimode squeezing Hamiltonian given by Eq.~(\ref{eq:HTMS}), where $G$ is the (as yet, undefined) adjacency matrix for a TMS graph.  Writing $U_\alpha$ as the Heisenberg matrix corresponding to $\exp(-i\alpha \mathcal H)$ gives
\begin{align}
\label{eq:U}
	U_\alpha =
	\begin{pmatrix}
		e^{\alpha G}	& 0			\\
		0			& e^{-\alpha G}	\\
	\end{pmatrix}
	\;.
\end{align}
A large (but finite) value of $\alpha$ is required for a useful CV cluster state.  Although previous work~\cite{Pfister2004,Bradley2005,Pfister2007} has emphasized uniformly weighted TMS graphs with no self-loops, at this point the only restriction we are going to place on $G$ is that it be {\em symmetric\/} and {\em full-rank\/}.  Experimental requirements will favor some $G$'s over others but, since any $G$ is in principle possible to implement~\cite{Lifshitz2005}, we will not impose any additional restrictions at this point.

With these requirements we can write $G$ as the difference of two positive semidefinite matrices that are mutually orthogonal.  By this we mean $G = G_+ - G_-$, where $G_\pm \ge 0$ and $G_\pm G_\mp = G_\mp G_\pm = 0$.  We write $G_\pm^\circ$ for the Moore-Penrose pseudoinverse of $G_\pm$, which (for symmetric matrices) is obtained by inverting all the nonzero eigenvalues of $G_\pm$.  Then, $G^{-1} = G_+^\circ - G_-^\circ$.  The projectors onto the positive and negative subspaces of $G$ are $P_\pm = G_\pm G_\pm^\circ = G_\pm^\circ G_\pm$.  Recalling Eq.~\eqref{eq:U}, we need both the positive and negative exponentials of $G$ in the limit of large $\alpha$.  In the positive (negative) case, such an operation will magnify all the positive (negative) eigenvalues of $G$ and zero out all of $G$'s negative (positive) eigenvalues.  To write this concisely, we start with the fact that $e^{-\alpha G_\pm} \to P_\mp$ for large $\alpha$, since all of the nonzero eigenvalues of $G_\pm$ get sent to zero (since $G_\pm \ge 0$) while the zero eigenvalues get raised to 1.  This gives
\begin{align}
\label{eq:expalphaG}
	 e^{\pm \alpha G} = e^{- \alpha G_\mp} e^{\alpha G_\pm} \to P_\pm e^{\alpha G_\pm} = G_\pm G_\pm^\circ e^{\alpha G_\pm}\;.
\end{align}
By suitably numbering nodes, the adjacency matrix for any bipartite graph can be written as
\begin{align}
\label{eq:AfromA0}
	A =
	\begin{pmatrix}
		0 		& A_0  \\
		A_0^T	& 0
	\end{pmatrix}
	\;,
\end{align}
where $A_0$ is $L \times (N-L)$.  Instead of using colors, we will label the first $L$ modes by~$+$ and the rest by~$-$ because the number of each will correspond to the number of positive and negative eigenvalues of $G$, respectively.  Recalling Eq.~\eqref{eq:CVCSsufficient}, we will use the phase-shift freedom in $T$ to rotate all of the $-$ modes by $-\pi/2$ and leave the others unchanged.  This gives
\begin{align}
\label{eq:minusAoneT}
	\begin{pmatrix}
		-A & \1
	\end{pmatrix}
	T
	&=
	\begin{pmatrix}
		0 &	-A_0	& \1	& 0  \\
		-A_0^T	 & 0		& 0	& \1
	\end{pmatrix}
	\begin{pmatrix}
		\1_+	& -\1_-  \\
		\1_-	& \1_+
	\end{pmatrix}
	\nonumber \\
	&=
	\begin{pmatrix}
		0 		& 0	& \1	& A_0  \\
		-A_0^T	& \1	& 0	& 0
	\end{pmatrix}	
	\;,
\end{align}
where $\1_\pm$ is the identity matrix on the $\pm$ modes and zero on $\mp$ modes and the identity blocks and zero blocks are sized appropriately, according to the dimensions of $A_0$.  Plugging Eqs.~\eqref{eq:U}, \eqref{eq:expalphaG}, and~\eqref{eq:minusAoneT} into Eq.~\eqref{eq:CVCSsufficient} gives the following sufficient condition for cluster state creation:
\begin{align}
\label{eq:minusAoneTUexpand}
	\begin{pmatrix}
		0 		& 0	& \1	& A_0  \\
		-A_0^T	& \1	& 0	& 0
	\end{pmatrix}
	\begin{pmatrix}
		G_+ G_+^\circ e^{\alpha G_+}	& 0						\\
		0						& G_- G_-^\circ e^{\alpha G_-}	\\
	\end{pmatrix}
	= 0
\end{align}
Keeping in mind that the first matrix is $N \times 2N$, while the second is $2N \times 2N$, this condition is fulfilled if 
\begin{align}
\label{eq:Gpmsimple}
	 \begin{pmatrix}
	 	-A_0^T	& \1
	\end{pmatrix}
	G_+ = 0 \qquad \text{and} \qquad
	\begin{pmatrix}
		\1		& A_0
	\end{pmatrix}
	G_- = 0\;.
\end{align}
These requirements are satisfied by choosing
\begin{align}
\label{eq:Gpmchoice}
	G_+ =
	\begin{pmatrix}
		\1	\\
		A_0^T
	\end{pmatrix}
	B
	\begin{pmatrix}
		\1	& A_0
	\end{pmatrix}
	, \;\;
	G_- =
	\begin{pmatrix}
		-A_0	\\
		\1
	\end{pmatrix}
	C
	\begin{pmatrix}
		-A_0^T	& \1
	\end{pmatrix}
	\;,
\end{align}
where $B,C > 0$ are arbitrary symmetric positive definite matrices.  This also illustrates our earlier point that labeling the sets of nodes as $+$ and $-$ reflected their connection to the number of eigenvalues of $G$ having each sign.  Thus, a CV cluster state with a bipartite adjacency matrix~$A$ satisfying Eq.~\eqref{eq:AfromA0} can be created with a TMS Hamiltonian of the form of Eq.~\eqref{eq:HTMS}, with
\begin{align}
\label{eq:Ggeneral}
	G &=
	\begin{pmatrix}
		\1		& -A_0	\\
		A_0^T	& \1
	\end{pmatrix}
	\begin{pmatrix}
		B	& 0	\\
		0	& -C
	\end{pmatrix}
	\begin{pmatrix}
		\1		& A_0	\\
		-A_0^T	& \1
	\end{pmatrix}
	\nonumber \\
	&=
	\begin{pmatrix}
		[B - A_0 C A_0^T]		& [B A_0 + A_0 C]	\\
		[C A_0^T + A_0^T B]		& [A_0^T B A_0 - C]
	\end{pmatrix}
	\;.
\end{align}
For a given $A$, this is the most general $G$ that satisfies Eq.~\eqref{eq:minusAoneTUexpand}, since $B$ and $C$ encompass all possible rotations of the eigenvectors and scalings of the eigenvalues that preserve the partitioning defined by Eq.~\eqref{eq:Gpmsimple}.  With $A$ fixed, the freedom to choose the $G$ that is easiest to implement experimentally is found solely within the choices of $B$ and $C$.

This is not the most general solution to the overarching problem, however.  There is no reason a priori that we should have a completely fixed $A$ for a given CV cluster state that we wish to create.  While all QND interactions in the original formulation~\cite{Menicucci2006} of CV cluster states for QC had the same strength, this is not necessary.  A weighted adjacency matrix~$A$ corresponds to variable-strength QND interactions for the edges of the graph.  This introduces squeezing and/or reversal ($q \to aq$, $p \to p/a$, where $a$ is the edge weight) to the Gaussian correction term that accumulates after each measurement.  While very low (or very high) weights would lead to difficulty resolving the quantum state after being heavily squeezed, for weights $\sim \pm1$, both theoretically and practically speaking, all of the quantum information is still preserved under single-mode measurements made on the cluster.  Allowing $A$ to be weighted gives additional degrees of freedom to the problem, allowing us even greater freedom in optimizing the experimental viability of the multimode squeezing Hamiltonian used to make our cluster state.

A corollary to this result is that any multimode squeezing Hamiltonian of the form of Eq.~\eqref{eq:HTMS} that has a full-rank $G$ generates some weighted bipartite CV cluster state (after appropriate single-mode phase shifts).  To see this, write $G$ in terms of its eigendecomposition $G = V\nu V^T$, where $\nu$ is a diagonal matrix of eigenvalues and $V$ is an orthogonal matrix.  Using elementary column operations, up to a possible renumbering of the output modes, we can always transform $V$ into the form of the first matrix in Eq.~\eqref{eq:Ggeneral}.  The target form always exists because it is the simultaneous column-reduced echelon form for the positive and negative subspaces of $G$ and $G$ is assumed to be full-rank.  These column operations, since they act separately on the two subspaces, can be represented by an invertible block-diagonal matrix $M$ acting from the right, such that $VM = \bigl( \begin{smallmatrix} \1 & -A_0 \\ A_0^T & \1 \end{smallmatrix} \bigr)$.  The transpose of this matrix, $M^T$, acting from the left, represents the same action as row operations on $V^T$.  With $M$ being invertible and block-diagonal, we can choose a $B,C>0$ such that $M^{-1} \nu M^{-T} = \bigl( \begin{smallmatrix} B & 0 \\ 0 & -C\end{smallmatrix} \bigr)$.  Thus, we can always write
\begin{align}
\label{eq:AfromG}
	G &= V M(M^{-1} \nu M^{-T}) M^T V^T \nonumber \\
	&=
	\begin{pmatrix}
		\1		& -A_0	\\
		A_0^T	& \1
	\end{pmatrix}
	\begin{pmatrix}
		B	& 0	\\
		0	& -C
	\end{pmatrix}
	\begin{pmatrix}
		\1		& A_0	\\
		-A_0^T	& \1
	\end{pmatrix}
\end{align}
for some particular $A_0$.  Comparing this with Eq.~\eqref{eq:Ggeneral}, we can immediately extract $A_0$ and use Eq.~\eqref{eq:AfromA0} to write~$A$ in terms of it.  This completes the proof.  We therefore also know that any multimode squeezing Hamiltonian generates a weighted bipartite CV cluster state (generally with a different graph~$A$) as long as the TMS adjacency matrix~$G$ is full-rank.

Intuitively, what's happening with this correspondence is that $\mathcal H$ from Eq.~\eqref{eq:HTMS} is used to squeeze the vacuum along $N$ joint quadratures (since $G$ is full-rank) with overall squeezing strength~$\alpha$.  
In general, these states are not CV cluster states because they do not satisfy Eq.~\eqref{eq:CVCSdef} for any choice of~$A$ in the large-$\alpha$ limit.  What we have shown is that by partitioning the resulting output modes into two groups (corresponding to the number of $\pm$ eigenvalues of $G$) and phase-shifting one of those groups by $-\pi/2$, we can always transform the output from the multimode squeezer into a CV cluster state, satisfying Eq.~\eqref{eq:CVCSdef} for some choice of $A$ as $\alpha$ becomes large.  Our derivation requires that $A$ be bipartite for this to work.

As an example, let $G$ be the complete graph on four nodes.  This generates a GHZ state~\cite{Pfister2004} whose quadrature operators satisfy $q_1 + q_2 + q_3 + q_4 \to 0$, $p_1 - p_2 \to 0$, $p_1 - p_3 \to 0$, and $p_1 - p_4 \to 0$ (and any linear combinations thereof).  Phase-shifting mode~1 (although any mode will do) by $-\pi/2$ means that now $-p_1 + q_2 + q_3 + q_4 \to 0$, $q_1 - p_2 \to 0$, $q_1 - p_3 \to 0$, and $q_1 - p_4 \to 0$, which satisfies Eq.~\eqref{eq:CVCSdef} with $A$ being the star graph on four nodes, with node~1 in the center.  This property generalizes: $G$ being the complete graph on $N$ nodes creates an $N$-mode GHZ state, which is equivalent to an $N$-mode star-graph CV cluster state after phase-shifting one of the output modes by $-\pi/2$.  The shifted mode becomes the central node in the star.  This mimics the case for qubits~\cite{Hein2006}, although the analogy is not exact since $G$ and $A$ represent different types of graphs (TMS and CV-cluster, respectively).

\begin{figure}[t*]
\begin{center}
\begin{tabular}{c}
\includegraphics[width= .85 \columnwidth]{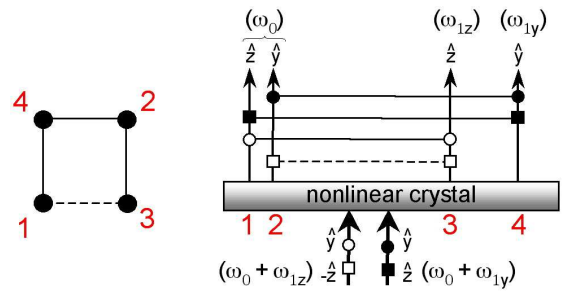}
\end{tabular}
\end{center}
\vspace{-.25in}
\caption{Experimental implementation of a square CV cluster state using a single OPO based on a periodically poled birefringent crystal, such as $\mathrm{KTiOPO_4}$ (see also Ref.~\cite{Pooser2005}). Left: the cluster graph ($A$) after a phase-shift of modes~3 and~4 by $-\pi/2$; dashed line denotes a negative weight; all magnitudes are $1/\sqrt{2}$. Right: the experimental proposal. Top arrows are OPO modes, bottom arrows are pumps, and all have polarization directions $\hat y$ or $\hat z$ along the crystal axes. Nonlinear interactions simultaneously phasematch (first letter is pump) $yyz$ (open circles), $yzy$ (filled circles), $zzz$ (open squares), and $zyy$ (filled squares). The OPO cavity resonance conditions and crystal birefringence ensure that no other OPO mode can be coupled to these four modes. Note the crucial importance of the $\pi$-shifted pump, $-\hat z$.}
\vspace{-.25in}
\label{fig:square}
\end{figure}

Star graphs are not universal for QC, however.  We'd like to achieve a procedure for generating a square-lattice (or other QC-universal) CV graph with a single OPO.  Such a graph is bipartite, so a corresponding $G$ can be constructed to create it and, in principle, can be quasi-phase-matched in a single photonic quasicrystal~\cite{Lifshitz2005}.  A significant step in this direction is the creation of a CV cluster state with a square graph from a single four-mode OPO: one can, indeed, show the remarkable result,
\begin{align}
\label{eq:square}
	A_0 = \frac {1} {\sqrt{2}} \begin{pmatrix}
		-1	& 1	\\
		1	& 1
	\end{pmatrix}
	\; \Longrightarrow \;
	G =
	\begin{pmatrix}
		0	& A_0 \\
		A_0^T& 0
	\end{pmatrix}
	= A
	\;.
\end{align}
Notice that $A$ is weighted so that one of the edges (sides of the square) has an opposite interaction sign to the three others and all have magnitude $1/\sqrt{2}$.  A (nonunique) generating $G$ is identical and immediately implementable using current technology, in fact, using the existing nonlinear crystal~\cite{Pooser2005} designed to produce the four-party CV GHZ state (Fig.~\ref{fig:square}).  Defining $\alpha$ as in Eq.~\eqref{eq:U}, the variance in each of the components of $\vec p - A \vec q$ for this state is $2e^{-2\alpha}$ units of vacuum noise.  Since these vanish as $\alpha \to \infty$, this is a valid square-graph CV cluster state.

In conclusion, we have shown that any continuous-variable cluster state with a bipartite graph can be generated from the application of a single multimode squeezing Hamiltonian.  We also have shown that all multimode squeezing Hamiltonians that have a full-rank two-mode squeezing adjacency matrix correspond to a weighted bipartite continuous-variable cluster state, generally with a different graph.  While as resource-efficient as the most efficient scheme currently known~\cite{vanLoock2006a}, these results are important for experiments because they provide a powerfully scalable means of generating continuous-variable cluster states using only one OPO and no beam-splitter network. 

We thank Gerard Milburn, Michael Nielsen, Tim Ralph, and Robert Jones for useful discussions and suggestions.  NCM acknowledges support from the U.S.\ Dept.~of Defense, STF from ONR Grant No.~N00014-07-1-0304, and OP from NSF Grant Nos.\ PHY-0555522 and CCF-0622100.


\end{document}